\documentclass[12pt]{article}
\usepackage{epsfig}
\begin{document}
\title{On the large-scale inhomogeneous Universe\\  
and the cosmological constant}
\author{D. PALLE \\
Zavod za teorijsku fiziku \\
Institut Rugjer Bo\v skovi\'c \\
Po\v st. Pret. 180, HR-10002 Zagreb, CROATIA}
\date{ }
\maketitle

\vspace{5 mm}
{\it
We study the large-scale inhomogeneity of the Universe
based on the averaging procedure of Buchert and Ehlers.
The generalized Dyer-Roeder equation
 for the angular diameter distance
of the inhomogeneous Universe is derived and solved for different
cosmological models. We make a comparison of certain
cosmic observables, such as the Hubble function, angular
diameter distance, cosmological correction factor of
homogeneous and inhomogeneous cosmological models, which are
crucial ingredients in galaxy number counts and gravitational
lenses.
}
\vspace{5 mm}

It seems that present cosmological data consisting of
supernova searches, structure formation, gravitational
lenses, etc. suggest a rather unexpected equation of
state for the simple homogeneous Friedmann-Lema$\hat{i}$tre (FL) 
Universe with $\Omega_{m}\simeq 0.3,\ \Omega_{\Lambda}
\simeq 0.7.$ 
However, one should be cautious even when performing analyses of
the cosmological data (for example, for supernova data see the reanalysis
in \cite{Rowan}). Further concern is the low
statistical confidence of all extracted observables.
Nevertheless, this phenomenological result
initiated a lot of speculative work on the
origin of the positive cosmological constant.

Contrary to current investigations, there is only
one derivation of the cosmological constant without
any fine tuning or any obscure assumptions
and this derivation 
 is based on the Einstein-Cartan gravity \cite{Palle1}.
When the Universe is frozen at zero temperature $T_{\gamma}=0$
(spacelike infinity) and if all nonrelativistic fermionic matter
is spinning matter, then owing to the same coupling between
spacetime curvature and mass density on the one side and
spacetime torsion and spin density of matter on the other
side, the following relations emerge: $ \Omega_{m}= 2,\ 
\Omega_{\Lambda}= -1 $ \cite{Palle1}.
It is important to strengthen that at $T_{\gamma}=0$ all 
spin densities of nonbaryonic and baryonic species act coherently
to the total spin.
The role of spin densities in the derivation of the 
primordial density fluctuation is described in Ref. \cite{Palle2}.
The existence of vorticity and acceleration \cite{Palle1,Palle2}
or nonvanishing shear \cite{Palle3} should be 
of fundamental importance for resolving certain cosmological
problems, but the cosmological
distance measures are not very much
affected by relatively small deviations from the FL model
\cite{Palle4}.
To conclude, let us notice that it is possible to 
construct a nonsingular and nonanomalous gauge theory
with heavy and light neutrinos as cold and hot dark matter
fermionic particles respectively, but without Higgs scalars and with
no asymptotoic freedom in QCD \cite{Palle5}.

In this paper we want to reconcile current measurements with
large-scale inhomogeneous cosmic models.
We assume that the Universe is inhomogeneous on large scales,
thus clumpiness should not be
characteristic only of small-scale
structures, and that the Earth is placed
in the region away from the centre of homogeneity.

Starting with the historical work of Lema$\hat{i}$tre, Tolman
and Bondi (LTB) on inhomogeneous models, 
one can account enormous activity in this field \cite{Krasinski}.
However, a recent attempt \cite{Iguchi} to exploit LTB models,
following perturbative calculations  
in Ref. \cite{Celerier}, was
not successful at large redshifts.

The cellular structure of the Universe with a power-law
distribution of matter of Ruffini et al \cite{Remo},
represents another approach to the problem of inhomogeneity. 

The work of Zalaletdinov \cite{Roustam} is the first 
complete and consistent treatment of the averaged
Einstein equations in an arbitrary Riemannian spacetime.
However, even if the vorticity and acceleration do not vanish,
they are small deviations (which we neglect in this paper) 
from the FL geometry, thus
we choose the following line element and spatially averaged
scalar quantites of Buchert and Ehlers \cite{Buchert}:

\begin{eqnarray*}
ds^{2} = dt^{2}-g_{ij}dX^{i}dX^{j},
\end{eqnarray*}
\begin{eqnarray*}
\langle \Psi(t,X^{i}) \rangle_{{\cal D}} \equiv
\frac{1}{V_{{\cal D}}}\int_{{\cal D}}d^{3}X J \Psi(t,X^{i}),
\end{eqnarray*}
\begin{eqnarray*}
\ V_{{\cal D}}(t)\equiv \int_{{\cal D}}J d^{3}X, 
\ J(t,X^{i})\equiv [det(g_{ij})]^{\frac{1}{2}}.
\end{eqnarray*}

Introducing natural definitions

\begin{eqnarray*}
a_{{\cal D}}(t) \equiv (\frac{V_{{\cal D}}(t)}
{V_{{\cal D}_{0}}})^{\frac{1}{3}},
\end{eqnarray*}
\begin{eqnarray*}
\langle \theta \rangle_{{\cal D}} =
\frac{\dot{V}_{{\cal D}}}{V_{{\cal D}}}=
3\frac{\dot{a}_{{\cal D}}}{a_{{\cal D}}},\ 
\dot{\psi}(t)\equiv \frac{d}{dt}\psi(t),
\end{eqnarray*}
\begin{eqnarray*}
M_{{\cal D}}=\int_{{\cal D}}d^{3}XJ\rho (t,X^{i}), \\
\langle \rho \rangle_{{\cal D}} = \frac{M_{{\cal D}}}
{V_{{\cal D}_{0}}a^{3}_{{\cal D}}}.
\end{eqnarray*}
one can easily derive Raychaudhuri and constraint equations
for averaged scalars
from Einstein and conservation equations \cite{Buchert,Kasai}

\begin{eqnarray}
3\frac{\ddot{a}_{{\cal D}}}{a_{{\cal D}}}
+4\pi G_{N}\frac{M_{{\cal D}}}{V_{{\cal D}_{0}}
a_{{\cal D}}^{3}} - \Lambda = Q_{{\cal D}},
\end{eqnarray}
\begin{eqnarray}
3(\frac{\dot{a}_{{\cal D}}}{a_{{\cal D}}})^{2}
-8\pi G_{N}\frac{M_{{\cal D}}}{V_{{\cal D}_{0}}
a_{{\cal D}}^{3}} + \frac{1}{2}\langle {{\cal R}} \rangle _{{\cal D}}
- \Lambda = -\frac{1}{2}Q_{{\cal D}}, 
\end{eqnarray}
\begin{eqnarray*}
Q_{{\cal D}} \equiv \frac{2}{3}\langle ( \theta
- \langle\theta\rangle _{{\cal D}})^{2}\rangle _{{\cal D}}
- 2\langle \sigma ^{2}\rangle _{{\cal D}},
\end{eqnarray*}
\begin{eqnarray*}
{\cal R}={\cal R}^{i}_{i},\ {\cal R}_{ij}\equiv spatial 
\ part\ of\ the\ Ricci\ tensor.
\end{eqnarray*}

In addition, we need a condition  of integrability
of these equations \cite{Buchert}:

\begin{eqnarray}
(a_{{\cal D}}^{6}Q_{{\cal D}})^{.}
+ a_{{\cal D}}^{4}(a_{{\cal D}}^{2}\langle {\cal R} \rangle
_{{\cal D}})^{.} = 0 .
\end{eqnarray}

Now we turn to the geometrical optics in order to
derive the wanted observables of the inhomogeneous cosmic
models.

Starting from the definition of the angular diameter
distance and the focusing equation \cite{Schneider}:

\begin{eqnarray*}
D \equiv (\frac{d A_{S}}{d \Omega_{O}})^{\frac{1}{2}},
\end{eqnarray*}
\begin{eqnarray}
\frac{d^{2}}{d w^{2}}A_{S}^{\frac{1}{2}} +
\frac{4\pi G_{N}}{H_{0}^{2}} (1+z)^{2} \rho A_{S}^{\frac{1}{2}} 
= 0 ,
\end{eqnarray}
\begin{eqnarray*}
w \equiv H_{0}\omega_{0}v,\ v=affine\ parameter.
\end{eqnarray*}

We take into account that
the shear of the congruence of photons vanishes for a spherically
symmetric source \cite{Sasaki}. Then, performing the Buchert-Ehlers spatial
averaging and neglecting the spacetime shear in the averaged equation
for the change of redshift with respect to the affine 
parameter:

\begin{eqnarray}
\frac{d z}{d w} = (1+z)^{3}\frac{\dot{a}_{{\cal D}}(z)}
{\dot{a}_{{\cal D}}(0)},
\end{eqnarray}

we arrive at the following equation for the averaged 
angular diameter distance:

\begin{eqnarray*}
\frac{d^{2}\langle D \rangle _{{\cal D}}}{d z^{2}}
(\frac{d z}{d w})^{2} + \frac{d \langle D \rangle _{{\cal D}}}
{d z}\frac{d^{2} z}{d w^{2}} + \frac{4\pi G_{N}}{H_{0}^{2}}
(1+z)^{2} \langle \rho D \rangle _{{\cal D}} = 0.
\end{eqnarray*}

Assuming the smooth space dependence of mass density and angular
diameter distance, one can factorize the last term of
the
preceding equation as
    
\begin{eqnarray*}
\langle \rho D \rangle _{{\cal D}} \simeq \rho (t,\bar{X^{i}})
\langle D \rangle _{{\cal D}} \simeq 
\frac{M_{{\cal D}}}{V_{{\cal D}}} \langle D \rangle _{{\cal D}},
\ \bar{X^{i}} \epsilon {\cal D},
\end{eqnarray*}

and we are left with the second-order 
generalized Dyer-Roeder equation with a
suitable density normalization for the cosmological models:

\begin{eqnarray}
\phi(z)\frac{d^{2} \tilde{D}(z)}{d z^{2}} +
\chi(z)\frac{d \tilde{D}(z)}{d z} +
\eta(z)\tilde{D}(z) = 0 ,
\end{eqnarray}
\begin{eqnarray*}
\tilde{D}\equiv \langle D \rangle,\ \phi(z)=(\frac{d z}{d w})^{2},
\ \chi(z)=\frac{d^{2} z}{d w^{2}},
\end{eqnarray*}
\begin{eqnarray*}
\frac{d^{2} z}{d w^{2}} = (\frac{d z}{d w})^{2}
(\frac{3}{1+z}+\frac{1}{\dot{a}_{{\cal D}}(z)}
\frac{d \dot{a}_{{\cal D}}(z)}{d z}), \\
\frac{d \dot{a}_{{\cal D}}}{d z} =
-\frac{1}{H(z)(1+z)}\ddot{a}_{{\cal D}}, \\
H(z)\equiv \frac{\dot{a}_{{\cal D}}(z)}{a_{{\cal D}}(z)},
\ a_{{\cal D}}(z)=\frac{1}{1+z},
\end{eqnarray*}
\begin{eqnarray*}
\eta(z) = \frac{3}{2}(1+z)^{5}\Omega_{m},\  
\Lambda = 3 H_{0}^{2} \Omega_{\Lambda}.
\end{eqnarray*}

Under adequate initial conditions \cite{Schneider}
(Adams-Bashforth integration method used)

\begin{eqnarray*}
for\ integration\ from\ z_{1}\ to\ 
z_{2},\ z_{2} > z_{1}, \\
\tilde{D}(z_{1},z_{1}) = 0,\ \frac{\tilde{D}(z_{1},z)}
{d z} (z=z_{1}) = \frac{1}{H(z_{1})}\frac{1}{1+z_{1}},
\end{eqnarray*}

we can calculate the averaged angular diameter distance for an arbitrary
cosmological model $\Omega_{m}+\Omega_{\Lambda}=1$.
Needless to say, for the vanishing backreaction and curvature terms one
recovers a homogeneous solution:

\begin{eqnarray*}
D(z) = \frac{1}{H_{0}(1+z)}\int ^{z}_{0}\frac
{d \zeta}{[\Omega_{m}(1+\zeta)^{3} + \Omega_{\Lambda}]
^{\frac{1}{2}}}.
\end{eqnarray*}

Numerical evaluations and comparisons are performed
in the following manner: (1) choose some functional
form for the backreaction Q(z), (2) evaluate curvature
term R(z) from the integrability condition,
(3) choose two inhomogeneous models with $\Omega_{m}=2$,
 $\Omega_{\Lambda}=-1$ [Einstein-Cartan (EC) model] and
$\Omega_{m}=1$ , $\Omega_{\Lambda}=0$ [Einstein-deSitter (EdS) model]
and fit parameters of the backreaction and curvature to
reach the input value for the Hubble constant $H_{0}=\dot{a}(z=0)/
a(z=0)$ and to get the best fit of the $\tilde{D}(z)$ of the inhomogeneous
models to the D(z) of the homogeneous model 
with \newline $\Omega_{m}=0.3$ , $\Omega_{\Lambda}=0.7$ for $z \leq 0.7$.

One can easily check that for any $0 > \alpha > -3$
(we omit index ${\cal D}$)

\begin{eqnarray*}
Q(z) = Q_{0} a^{\alpha}(z),
\end{eqnarray*}
\begin{eqnarray*}
\langle {\cal R} \rangle _{{\cal D}} \equiv
R(t) = -a^{-2}(t)\int ^{t}_{t_{0}} d \tau a^{-4}(\tau)
\frac{d}{d \tau} [a^{6}(\tau)Q(\tau)],
\end{eqnarray*}
\begin{eqnarray*}
R(z)=-\frac{\alpha+6}{\alpha+2}[Q(z)-(\frac{a(z_{0})}{a(z)})^{2}
Q(z_{0})],\ \alpha\neq -2, \\
R(z)=4Q(z)\ln\frac{1+z}{1+z_{0}},\ \alpha=-2 ,
\end{eqnarray*}

the curvature term power-law coefficient is within the same interval 
as that for the backreaction. Thus, this simple functional
form garantees the homogeneity at very large scales
$\mid Q(z) \mid , \mid R(z) \mid \ll H_{0}^{2}\Omega_{m}
a_{{\cal D}}(z)^{-3},
\ z >> 1$(it could also be worth including the  Gaussian damping, if necessary).

For definite numerical comparisons of the three cosmological
models we use this Ansatz:

\begin{eqnarray*}
Q(z)=Q_{0}[1+c_{0}a^{-1}(z)], \hspace{50 mm}\\
R(z)=Q_{0}(1+z)^{2}[-3(\frac{1}{(1+z)^{2}}
-\frac{1}{(1+z_{0})^{2}})+5c_{0}(\frac{1}{1+z}
-\frac{1}{1+z_{0}})].
\end{eqnarray*}

In order to match the input Hubble parameter at present time $H_{0}$
, $z_{0}$ has to fulfil the following equation:

\begin{eqnarray*}
Q(0)+R(0)=0 \hspace{45 mm} \\
\Rightarrow\ \  2(-1+3c_{0})(1+z_{0})^{2}-5c_{0}(1+z_{0})+3=0.
\end{eqnarray*}

Searching for the best fit values of $c_{0}$ and $Q_{0}$, we
find 
\begin{eqnarray*}
H_{0} \equiv h_{0}\ u,\ u\equiv 100 km s^{-1} Mpc^{-1}, \hspace{40 mm}\\
(EC):\ h_{0}=0.6,\ Q_{0}=3.789\ u^{2},\ z_{0}=\sqrt{6}/2-1,\ c_{0}=0, \hspace{8 mm} \\
(EdS):\ h_{0}=0.6,\ Q_{0}=1.2632\ u^{2},\ z_{0}=\sqrt{6}/2-1,\ c_{0}=0, \\
(Hom):\ h_{0}=0.6,\ \Omega_{m}=0.3,\ \Omega_{\Lambda}=0.7,
\ Q_{0}=0. \hspace{15 mm}
\end{eqnarray*}

For $z \leq 0.5$, the angular diameter distances differ less than
$5 \%$.
Figs. 1 and 2 show the Hubble flow and 
the angular diameter distance for the three models, while
Table 1 presents the cosmological correction terms for              
the difference in light travel time between 
gravitational lens images \cite{Refsdal}:

\begin{eqnarray*}
H_{0} \bigtriangleup t= T f(\theta_{1,Obs},\theta_{2,Obs},
z_{d},z_{s}),
\end{eqnarray*}
\begin{eqnarray*}
T = H_{0}\frac{\tilde{D}^{2}(0,z_{d})}{\tilde{D}}
(1+z_{d})\frac{z_{s}-z_{d}}{z_{s}z_{d}},\\
\tilde{D}=\frac{\tilde{D}(z_{d},z_{s})\tilde{D}(0,z_{d})}
{\tilde{D}(0,z_{s})}.
\end{eqnarray*}

\begin{figure}
\epsfig{figure=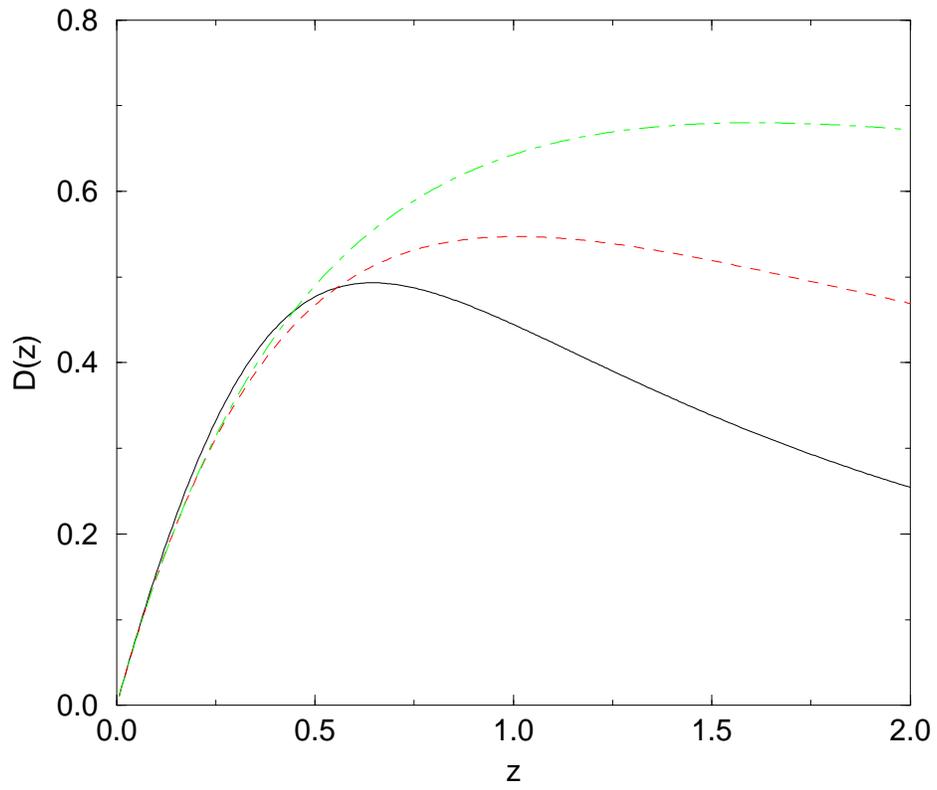, height=120 mm, width=140 mm}
\caption{ 
 Solid, dashed and dot-dashed lines 
denote angular diameter distance D(z) in [$u^{-1}$] for (EC), (EdS) and (Hom) models, 
respectively.}
\end{figure}

\begin{figure}
\epsfig{figure=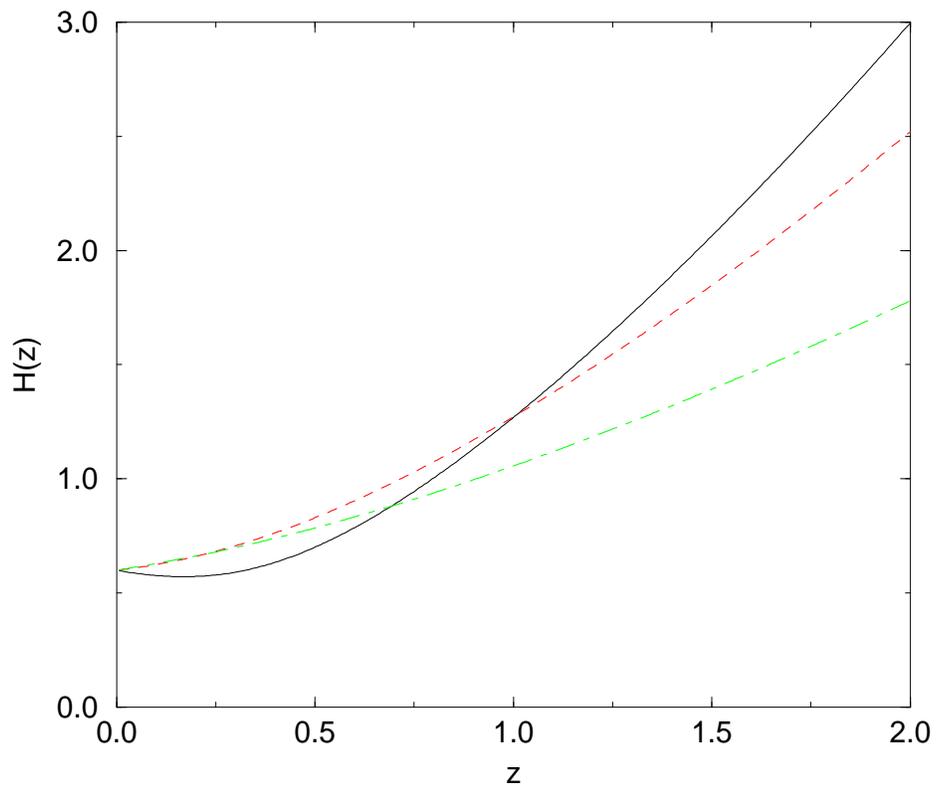, height=120 mm, width=140 mm}
\caption{
 Solid, dashed and dot-dashed lines 
 denote Hubble function H(z) in [u] for (EC), (EdS) and (Hom) models, respectively.}
\end{figure}

\vspace{5 mm}

\begin{tabular}{|c||c|c|c|c|}
  \hline
   &$z_{d}=0.5,\ z_{s}=1$&$z_{d}=0.5,\ z_{s}=1.5$&$z_{d}=1,\ z_{s}=1.5$
&$z_{d}=1,\ z_{s}=2$\\
  \hline
  \hline
 T(EC)& 0.7550 & 0.6686 & 0.4977 & 0.4226 \\
 T(EdS)& 0.9559 & 0.9350 & 0.8775 & 0.8471 \\
 T(Hom)& 1.8543 & 2.1098 & 4.7948 & 8.9155 \\
  \hline
\end{tabular}

\vspace{5 mm}

\noindent
The galaxy number count depends essentially on the
cosmological model through the comoving volume:
\begin{eqnarray*}
\frac{d V(z)}{d z} \propto \tilde{D}(z)^{2}.
\end{eqnarray*}

We conclude with two observations: (1) For the large-scale
inhomogeneity of the Universe to be established one needs 
clear indications that none of homogeneous models can 
simultaneously fit and explain all data in cosmography,
 structure evolution, gravitational lenses, etc. for
all redshifts.
At present it is difficult to
make any conclusion what is the source of some
recently found disparities in gravitational lenses \cite{Kochanek}
or galaxy evolution \cite{Sasseen}.
 (2) If the inhomogeneity is established, then one could
intend to model a backreaction function for the global fit
of data, and ultimately attempt to derive it from 
the evolution of multicomponent imperfect fluid
with photons, baryonic and nonbaryonic matter, with
a necessary knowledge of all relevant masses and couplings.
\newline

\hspace{58 mm}* * *  \newline
This work was supported by the Ministry of Science and Technology
of the Republic of Croatia under Contract No. 00980103.

\end{document}